\documentclass[floatfix,showkeys,twocolumn,citeautoscript,showpacs,amsmath,amssymb,epsf,aps]{revtex4}
\usepackage{dcolumn}
\usepackage{epsfig}

\usepackage{graphicx}
\usepackage{longtable}
\usepackage{dcolumn}
\usepackage{bm}

\def\BNF{B$_{24}$N$_{24}$ }

\def\BNE{B$_{28}$N$_{28}$ }
\def\BNT{B$_{32}$N$_{32}$ }
\def\BNS{B$_{36}$N$_{36}$ }
\def\BNR{B$_{96}$N$_{96}$ }

\def\CSX{C$_{60}$}
\def\CTFT{C$_{240}$}
\def\CNSXT{C$_{960}$}

\def\CTOS0{C$_{2160}$}
\def\AlNF{Al$_{24}$N$_{24}$ }
\def\AlNE{Al$_{28}$N$_{28}$ }
\def\AlNFE{Al$_{48}$N$_{48}$ }
\def\AlNT{Al$_{32}$N$_{32}$ }
\def\AlNS{Al$_{36}$N$_{36}$ }
\def\AlNR{Al$_{96}$N$_{96}$ }
\def\Xa{X$\alpha$}
\def\af{$\alpha$}

\def\trip_b{$^3$B$_1$ }
\def\sing_a{$^1$A$_1$ }
\def\af{$\alpha$ }

\def\Xa{X$\alpha$ }
\def\rhup{\rho_{\uparrow}}
\def\rhdn{\rho_{\downarrow}}
\def\vrp{\vec{r'}}
\def\vr{\vec{r}}
\def\bo{\overline{\rho}^{\frac{1}{3}}}
\def\ro{\overline{\rho}}
\def\bt{\overline{\rho}^{\frac{2}{3}}}
\def\ss{\sigma}
\def\bo{\overline{\rho}^{\frac{1}{3}}}
\def\ro{\overline{\rho}}
\def\bt{\overline{\rho}^{\frac{2}{3}}}
\def\gbo{\overline{g}^{\frac{1}{3}}}

\def\gbt{\overline{g}^{\frac{2}{3}}}
\def\ss{\sigma}

\begin{document}
\preprint{USC/002}

\title{ Fully analytic implementation of density functional theory for efficient calculations on large molecules}


\author{Rajendra R. Zope}
\email{rzope@alchemy.nrl.navy.mil}
\affiliation{Department of Chemistry, George Washington University, Washington DC, 20052, USA}
\affiliation{Department of Physics, University of Texas at El Paso, El Paso, Texas 79959, USA}

\author{Brett I. Dunlap}
\email{dunlap@nrl.navy.mil}
\affiliation{$^3$ Code 6189, Theoretical Chemistry Section, US Naval Research Laboratory,
Washington, DC 20375, USA}

\date{\today}

\begin{abstract}

  Fullerene like cages and naonotubes of carbon and other  inorganic materials
are currently under intense
study due to their possible technological applications. First principle
simulations of these materials are computationally challenging due to 
large number of atoms. We have recently developed a fast, variational
and fully analytic density functional theory (ADFT) based model that allows 
study of
systems larger than those that can be studied using existing density 
functional models.  Using polarized Gaussian basis sets (6-311G**)
and ADFT, we optimize geometries of  large fullerenes, 
fullerene-like  cages and nanotubes of    carbon, boron nitride,
and aluminum nitride containing more than two thousand atoms. The 
calculation of \CTOS0\, using nearly 39000 orbital basis functions is 
the largest calculation on any isolated molecule reported to-date 
at this level of theory, and it includes full geometry optimization.
  The electronic structure related properties 
of  the inorganic cages and other carbon fullerenes have been studied.

\end{abstract}

\pacs{ }


\maketitle

               Computer simulations are playing increasingly important role  
in our understanding about materials. Generally, the choice of 
computational models that are employed 
in studying the properties of materials depend on the property of interest and the 
length scale or the size of the system\cite{Other}. The latter is the most important 
factor in the selection 
of appropriate level of theory.  Our interest is in the electronic and structural 
properties of large carbon fullerenes and fullerene like cages of aluminum and boron 
nitride containing a few hundred atoms. At these sizes, the current toolbox 
of methods that are available include semiempirical quantum mechanical models such 
as ZINDO\cite{ZINDO}, PM3\cite{PM3} methods or tight binding approaches\cite{tight_binding}.
More accurate description of electronic properties require use of more involved 
methods such as density functional (DF) based models\cite{HK,KS65}.
The traditional quantum chemical beyond Hartree-Fock methods or quantum Monte Carlo 
are, in general, more accurate than the DF models. However, they are suitable for 
systems  with a few tens of atoms.
At present, the applicability of DF models is restricted to two to three hundred atoms
depending on the schemes used to approximate kinetic and exchange energy functionals, 
the basis sets used to expand the Kohn-Sham orbitals, the treatment of core electrons
(use and quality of pseudopotentials), and the type of atoms in the system.  We are working 
towards development of fully analytic implementation of density functional 
theory(ADFT)\cite{Dunlap03,ZD_JCTC}. 
The computationally efficient ADFT and efficient use of the available point group
symmetry of molecules allow us to optimize large inorganic and carbon fullerenes
containing more than two thousand atoms\cite{Dunlap05,ZD06}.

          The ADFT uses analytic atom-centered, localized Gaussian basis sets. These basis sets are used to 
to expand the molecular (Kohn-Sham) orbitals and the one body effective (Kohn-Sham) potential 
using variational and robust fitting methodology\cite{Dunlap79,BID89}. The exchange-correlation 
part of  Kohn-Sham potential is obtained using the functional form that 
is based on Slater's exchange functional\cite{Slater51}. For this reason, the analytic implementation 
is also called as Slater-Roothaan (SR) method\cite{Dunlap03}. The SR method allows an arbitrary 
scaling of the exchange potential around each type of atoms in the heteroatomic
systems. These scaling factors can be used to parametrize the SR method. Using 
a suitable choice of these scaling parameters, accurate total and atomization 
energies that are comparable to some of the most sophisticated density functional 
models can be obtained\cite{ZD_JCTC,ZD_PRB05,ZD_JCP06}.  In the following section we describe the analytic implementation 
of the density functional model  and  the details of the SR method.

\subsection{Analytic formulation of the G\`asp\`ar-Kohn-Sham-Slater density functional model}
In the Hohenberg-Kohn-Sham formulation of the density functional theory\cite{HK,KS65}
the total electronic 
energy of system containing $N$ electrons and $M$ nuclei is given by 
\begin{equation}
E^{HKS} [\rho]  = \sum_i^N <\phi_i| f_1| \phi_i> + E_{ee} + 
            E_{xc}\big [ \rhup, \rhdn \big] \label{eq:I}
\end{equation}
 where,  the first term contains the kinetic energy operator and the nuclear attractive potential due to
the $M$ nuclei,
\begin{equation}
f_1 = -\frac{\nabla^2}{2} - \sum_A^M \frac{Z_A}{\vert\vr - \vec{R_A}\vert}.
\end{equation}
The  second term in Eq. (\ref{eq:I}) represents the classical Coulomb interaction energy
of electrons while the last term  is the exchange-correlation energy that 
represents  contributions that are quantal in origin.
The Eq. (\ref{eq:I}) is an exact expression for the total energy but practical application 
require approximation to the $E_{xc}$. Over the years numerous parameterization of different accuracy 
and complexity have been devised and are  available in literature to model $E_{xc}$. 
Most of them however have complex functional 
form making use of  numerical grids necessary in implementations of the DF models.
Number of groups have developed numerical integration methods for computation 
of integrals over the exchange-correlation contributions\cite{Grids}. 
Today practically all implementations of the DF models use  
numerical grids to compute the contribution 
to the total energy and matrix elements from the exchange-correlation terms. This is true 
even if analytic basis sets such as Gaussian are used to express KS orbitals. However, 
it turns out that  if one models the $E_{xc}$ according to G\`asp\`ar-Kohn-Sham-Slater  
then the contribution to total energy from this term can also be calculated analytically
using the  Gaussian basis sets and variational methodology\cite{BID89,Cook95,Cook97}.  

 The  G\`asp\`ar-Kohn-Sham-Slater  (GKS) exchange energy functional is given by 
\begin{equation}
E_{xc} [\rhup, \rhdn]  =  - \frac{9}{8} \alpha \Big ( \frac{6}{\pi} \Big )^{1/3} \int d^3r 
            \Big [ \rhup^{\frac{4}{3}} (\vr) +  \rhdn^{\frac{4}{3}} (\vr) \Big]. \label{eq:Exc}
\end{equation}
  where, $\alpha = 2/3$ is the  G\`asp\`ar-Kohn-Sham value and $\alpha=1$  is the Slater's value.
In order to calculate  $E_{xc}$ analytically the one-third and two-third powers of the electron density 
are expanded in 
Gaussian basis sets: 
\begin{eqnarray}
            \rho^{\frac{1}{3}} (\vr) \approx \bo =  \sum_i e_i  E_i  \\
                                              \label{G2}
            \rho^{\frac{2}{3}} (\vr) \approx \bt =   \sum_i f_i  F_i .
                                              \label{G3}
\end{eqnarray}

 Here, $\{E_i\}$ and $\{F_i\}$ are independent Gaussian basis functions, while $e_i$ and $ f_i$ are expansion coefficients.
The exchange energy is then given by\cite{BID89,Cook95,Cook97}
\begin{equation}
E_{xc}   =  C_{\alpha} \Bigl [ \frac{4}{3} \langle \rho \,  \bo \rangle - \frac{
2}{3} \langle \bo  \,
     \bo  \, \bt \rangle
    + \frac{1}{3} \langle \bt \, \bt \rangle
 \Bigr ] ,
             \label{Eq:ex}
\end{equation}
 where $C_{\alpha} = - {9} \alpha \Big ( \frac{3}{\pi} \Big )^{1/3} .$
  Thus using four LCGO basis sets (one for orbital expansion and three for the fitting the Kohn-Sham potential) the 
total energy is calculated analytically.

    Similarly, to compute the Coulomb energy,

\begin{equation}
  E_{ee} = \langle  \rho\vert\vert\rho \rangle  =
    \frac{1}{2} 
 \int  \int \frac{\rho(\vr)\rho(\vrp)}{\vert\vr-\vrp\vert} d^3r \, d^3r',
\end{equation}
we use the first robust and variational fitting  methodology  and express  
the  charge density as a fit to a set of Gaussian
functions,
\begin{equation}
\rho(\vr) \approx \overline{\rho}(\vr) = \sum_i d_i G_i(\vr).
               \label{Eq:D}
\end{equation}
Here, $\ro (\vr)$ is the fitted density, d$_i$ is the expansion coefficient of the charge 
density Gaussian basis-function G$_i$. The elimination of the first order error in total energy due to the fit
leads to the unique robust expression for the self-Coulomb energy\cite{Dunlap79}. 
The LCAO orbital coefficients and the vectors {\bf d}, {\bf e}, and {\bf f} are found by constrained
variation.
It is easy  
to  obtain contribution from  the first term in Eq. (\ref{eq:I}) in analytic fashion. Thus, in ADFT four sets of 
Gaussian basis are required: one for KS orbitals, three for the KS effective potential. This 
methodology was successfully 
implemented by Werepentski and Cook who demonstrated that noise-free forces and smooth potential energy 
can be obtained using a {\em fully  analytic} (grid-free) implementation\cite{Cook95,Cook97}.

\subsection{Slater-Roothaan method}
       
   While the  above analytic implementation is computationally efficient 
its performance for the atomization of molecules is limited due to the limitation 
of the functional form adopted. We have tested its performance by computing 
atomization energies of a set of 56 molecules from the G2 dataset. For $\alpha = 2/3$,
the mean absolute error in atomization of 56 molecules is about 16 kcal/mol. This 
can be improved to 12 kcal/mol by allowing value of $\alpha$ to change\cite{ZD_CPL04}.
Thus the 
$\alpha$  in Eq. (\ref{Eq:ex}) can be viewed as a scaling parameter that scales 
exchange potential. The above model then can be modified so that each type 
of atom in the heteroatomic system has its own value of scaling parameter. This led to the 
development of the Slater-Roothaan (SR) method\cite{Dunlap03}. Apart from the advantage that
the calculations can be performed in complete analytic fashion, it also allows  
molecules to dissociate correctly in atomized limit\cite{Slater72}. The exchange energy in the SR 
method has following form\cite{Dunlap03,ZD_JCP06}:
\begin{eqnarray}
  E^{SR} &  = & \sum_i <\phi_i| f_1| \phi_i> + 2 \langle \rho \vert\vert\ro\rangle
                - \langle \ro\vert\vert\ro\rangle    \nonumber \\
          & & \,\, - \sum_{\sigma = \uparrow, \downarrow}
             C_x \Bigl   [
            \frac{4}{3} \langle g_{\ss}  \, \gbo_{\!\ss} \rangle
           - \frac{2}{3} \langle {\gbo}_{\!\ss} \,  {\gbo}_{\!\ss}  \, {\gbt}_{\!\ss} \rangle
  \nonumber \\
         & & \,\, \, +  \,\frac{1}{3} \langle {\gbt}_{\!\ss} \,  \, {\gbt}_{\!\ss} \rangle 
    \Bigr ].
             \label{G4}
\end{eqnarray}
  Here, $C_x = C_{\alpha}/\alpha$; the partitioned $3/4$ power of the exchange 
energy density,
\begin{equation}
  g_{\ss} (\vr) = \sum_{ij} \alpha(i) \, \alpha(j) \, D_{ij}^{\ss} (\vr),
\end{equation}
where $D_{ij}^{\ss} (\vr)$ is the diagonal part of the spin density matrix multiplied by 
the partitioning function,
\begin{equation}
   \alpha(i) =  \alpha_i^{3/8}
\end{equation}
which contains $\alpha_i$, the \af in the \Xa model for the atom on which the atomic orbital $i$ is
centered.  
The fits  to powers of $g_{\sigma}$ are obtained 
variationally from Eq. (\ref{G4}).

\section{Computational details}

  As noted earlier the analytic SR method requires four Gaussian basis sets.  One for the orbital 
expansion 
and others to fit different powers of electron density, which we obtain from 
literature.
%
 We choose Pople's triple-$\zeta$ (TZ) 6-311G** basis\cite{O1,O2} and the DGauss\cite{AW91} valence
double-$\zeta$  (DZ) basis set\cite{GSAW92} called DZVP2 for orbitals basis sets.  
The {\sl s-}type fitting bases are obtained by scaling and uncontracting 
the {\sl s} part of 
the orbital basis.  The scaling factors are 2 for the density, 
$\frac{2}{3}$ for $\bo$ \,\, and $\frac{4}{3}$ for $\bt$.  These  scaled bases are used for 
all {\sl s-}type fitting bases.
Ahlrichs' group has generated a RI-J basis for fitting the charge density of a valence 
triple-$\zeta$ orbital basis set used in the {\sc Turbomole}
program \cite{EWTR97}.  The non-$s$ parts of Ahlrich's fitting bases are used in 
combination with 
the 6-311G**  orbital basis sets.  We use this combination of 
basis sets (6-311G**/RIJ ) for boron nitride cages and carbon fullerenes. 
In combination with DZVP2 orbital 
basis, we use the {\sl pd} part of the  A2 charge density fitting basis. The 
combination  DZVP2/A2 is used for   studying aluminum nitride cages.
The geometries of molecules were optimized using the Broyden-Fletcher-Goldfarb-Shanno (BFGS) 
algorithm\cite{BFGS5}.
The forces on atoms are rapidly computed non-recursively using the 4-j generalized Gaunt 
coefficients \cite{Dunlap02, Dunlap05}.   The atomic energies are obtained in the highest 
symmetry for which the self-consistent solutions have integral occupation numbers.
The atomization energy is computed from the total energy difference of optimized 
molecule and its constituent atoms.  


\begin{figure}
\epsfig{file=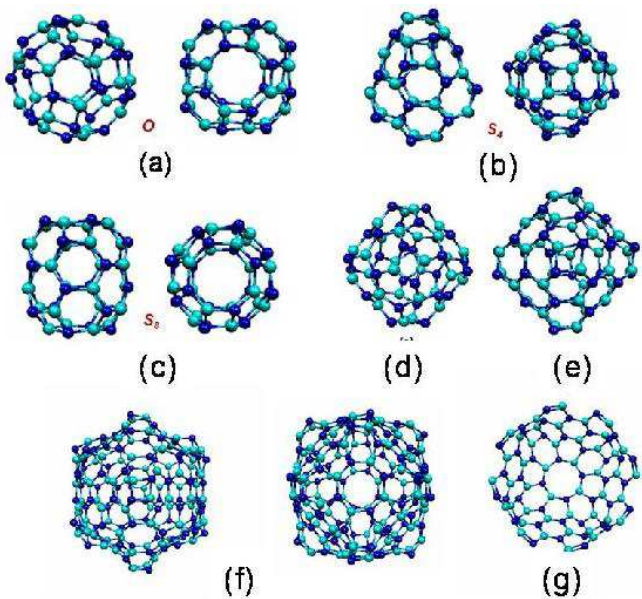,width=\linewidth,clip=true}
\caption{\label{fig1} The optimized BN cages: (a) Two views of the octahedral \BNF cage, (b) Two views of 
the S$_4$  \BNF\, cage, (c)  S$_8$ \BNF\, cage, (d) \BNE\, cage of $T$ symmetry, (e)  \BNS\, cage of 
T$_d$ symmetry, (f) octahedral \BNR\, cage, and (g) hemispherical  cap of (8,8) BN nanotube based 
on half of \BNR.}
\end{figure}

\begin{figure}
\epsfig{file=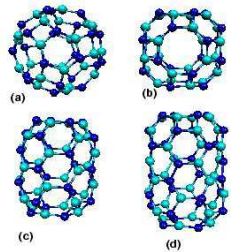,width=\linewidth,clip=true}
\caption{\label{fig2} The optimized structures of capped BN nanotubes. (a) and (b) are
two different views of the \BNF cage: (a) along the C$_3$ axis,  (b) along C$_{4}$ axis.  (c) \BNE ({\sl C$_{4h}$}) cage 
obtained by adding a ring of 8 alternate B and N atoms, (d) tubular \BNT ({\sl S$_8$}) cage by inserting
two rings of 8 alternate B and N atoms (see text for more details). }
\end{figure}
%
\begin{table}
\caption{The calculated values of binding energy per AlN pair, the energy gap between the highest 
occupied molecular orbital and the lowest unoccupied molecular orbital, the vertical ionization
potential (VIP), the electron affinity, and the energy gap obtained from the $\Delta SCF$  
calculation for the optimized AlN cages.  Last row gives range of values for the same 
set of BN clusters.  All energies are in eV.}
\label{table:BE}
\begin{tabular}{llrlrll}
\hline
             & Symmetry  &  BE     &    GAP        & VIP       &   VEA  & $\Delta ~SCF$ \\
\hline         
\AlNF        &  \sl{O}      &   10.24       &  2.97   & 7.05   &  1.46  &   5.59        \\
\AlNF       &  \sl{S$_4$}   &   10.34       &  2.47   & 6.84   &  1.72  &   5.12        \\
\AlNF       &  \sl{S$_8$}   &   10.34       &  2.63   & 6.79   &  1.58  &   5.21        \\
\AlNE        & \sl{C$_{4h}$}&   10.42       &  2.74   & 6.81   &  1.59  &   5.22        \\
\AlNE        & \sl{T}       &   10.45       &  2.67   & 6.84   &  1.69  &   5.15        \\
\AlNT        & \sl {S$_8$}  &   10.49       &  2.79   & 6.77   &  1.61  &   5.16        \\
\AlNS        & \sl {T$_d$}  &   10.54       &  2.70   & 6.73   &  1.76  &   4.95        \\
\AlNFE       & \sl {S$_d$}  &   11.09       &  2.81   & 6.56   &  1.76  &   4.8        \\
\AlNR        & \sl {O}      &   10.48       &  2.18   & 6.15   &  2.34  &   3.8  \\
\hline
 BN range    &               &  ~15         &  ~4-5   & ~7-9    &  ~0    & ~7-9   \\
\hline
\end{tabular}
\end{table}


\section{Boron and aluminum  nitride cages}
           The discovery of  carbon fullerene \CSX, followed by discovery of higher fullerenes and 
carbon nanotubes has led to intense search for  hollow cage-like and tube-like structures of other
materials.  In this search, boron nitride  (BN) is probably the second most studied material after carbon.
A number of groups have reported observation of BN nanotubes as well as cage like 
structures\cite{Oku03,Stephan98,Goldberg98,Oku00}. Particular relevant to this article are 
the series of experiments by Oku and coworkers 
in which they detected BN clusters in mass spectrum\cite{Oku00,Oku03}. These authors have proposed a 
number of cage like structures
for the BN clusters detected in mass spectrum. Here, we report the electronic structure 
of these cages and their aluminum nitride (AlN) analogues. We note that while 
of the BN cages has been reported, no cage like structure of AlN have not yet been observed 
although observations of the AlN nanotubes have been 
recently noted\cite{Tondare02,Bala04,N_Bala04,Expt_JACS}. 

                       The optimized cage structures of BN are given in Fig. {\ref{fig1}. Also,
given are the symmetries 
of these cage structures. All these cage structures have been found to be energetically stable with binding 
energy (BE) of about 14-16 eV per pair of BN.  Notable amongst these is the octahedral \BNF cage that 
was proposed by Oku and coworkers as a candidate structure 
for one of the most abundant cluster in the mass spectrum.  This cage 
is perfectly round and like in \CSX\, fullerene where each carbon atom is equivalent to all other 
carbon atoms, a pair of BN in this cluster is equivalent to other pairs in 
the cluster. It is to be noted that the exact analogue of carbon fullerene is not possible for alternate
boron nitride cages. The presence of pentagonal rings in carbon fullerenes do not permit full alternation 
of B and N atoms. Thus even membered rings are necessary to make alternate fullerenes close. 
The octahedral round \BNF cage contains six square and six octagons. This structure can be used to form 
caps for (4,4) BN nanotubes\cite{ZD_BN04}. However, unlike \CSX\, fullerenes, the round \BNF octahedral cage is not 
energetically special. The other alternate \BNF cages with symmetry S4 and S8 are energetically 
nearly degenerate with octahedral cage\cite{ZBPD04}. So it is not clear that which structure is 
likely to be observed in the experiment.

                    The C$_{4h}$ \BNE  cage [Fig.~\ref{fig2}(c)] can be  generated from the base \BNF 
cage by  cutting the latter into two halves after orienting it along the C$_4$ axis,
then inserting a ring of eight alternate B and N atoms perpendicular to the axis,
i.e. horizontally, and then rotating the top half by an eighth of a revolution.
The resultant cage  contains 8 inequivalent atoms and has 
{\sl C$_{4h}$} symmetry. If  two 
rings of  four alternate BN pair are inserted instead of one  then the resultant \BNT  cage is a 
tubular structure with {\sl S$_8$}  symmetry.   The binding energy systematically increases 
by going from \BNF to the  \BNE cage (0.26 eV per BN pair) and from \BNE to \BNT cage
(0.06 eV/BN pair).  The successive additions of alternate BN rings 
energetically stabilizes the BN tubular cages and results in  (4,4) BN nanotube with round 
caps that are based on  octahedral \BNF cage. Note that same (4,4) tube can also be 
generated by starting with S8 \BNF cage structure. The hemispherical caps of (4,4) tube 
based on octahedral \BNF that we have proposed have also been observed in 
a molecular dynamics study of the growth mechanism of BN nanotubes\cite{XVCC98}.

 The  \BNF can be enlarged by adding hexagons.  This leads to another round cage \BNR of octahedral 
symmetry.  The optimized \BNR cage is shown in Fig. ~\ref{fig1} (f).  Energetically the \BNR cage 
is more stable than the \BNF cage.  It is clearly different, from \BNF, in that while being
mostly round, its twelve squares stick out significantly, like the detonators of a sea mine.  
Its halves can form a  round cap for the (8,8)  BN nanotube (See Fig. ~\ref{fig1} (g)).


\begin{figure}
\epsfig{file=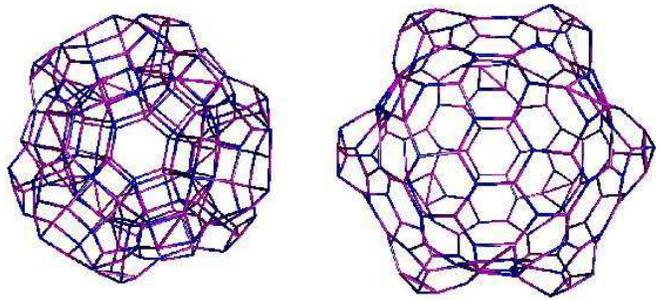,width=\linewidth,clip=true}
\caption{\label{fig:AlN}  Optimized octahedral \AlNR cages: (a) Two shell onion-like octahedral cage with  \AlNF\, 
cage at its interior \AlNR-I, (b)  Fullerene-like cage \AlNR-II.}
\end{figure}


      We have reoptimized  the BN cages by replacing boron by aluminum. We would like to point 
out that unlike BN cages which are experimentally observed, the AlN cages studied here are 
predictions.  The optimized cage structures in AlN are similar to those of BN except that they are 
larger in size due to the larger AlN bond distance than that of BN. The exception to this trend 
is the \AlNR cluster (See Fig. \ref{fig:AlN}).  
The optimization  of \AlNR structure starting from \BNR cluster results 
in formation of double shell onion like structure. This onion cluster has \AlNF cage at its 
core\cite{ZD_AlN}. On the other hand, if one scales the \BNS to account for larger AlN bond length and then 
replace B by Al and optimize then one does get  fullerene like hollow cage with squares sticking 
out. This cage will be refereed to as {\AlNR}-II.
We find that all AlN cages are energetically stable with binding energy of about 
10-11 eV per pair of AlN. However, the binding energy of AlN cages is less that BN cages, which 
have binding energy of 14-16 eV per BN pair. Similarly, the  ionization potential (IP) and the 
energy gap between the energy eigenvalues of highest occupied molecular orbital (HOMO) and 
the lowest unoccupied molecular orbital is smaller in AlN cages. The vertical IP of BN cages
is in the range 7-9 eV while it is about 6-7 eV in AlN cages. Due to the large HOMO-LUMO gap 
in the BN cages, the BN cages do not like to bind an extra electron. Consequently, the  electron
affinity of boron nitride cages is practically zero within our model. The AlN cages have electron 
affinity of about 1-2 eV. We have summarized the electronic structure data of AlN cages 
in Table I. In the last row of the same table contains the range of values for the BN cages.
The HOMO-LUMO gap calculated  by the so called $\Delta ~SCF$  
method  in which the first ionization potential is subtracted from the first 
electron affinity is given in the last row of the table.


\begin{figure}
\epsfig{file=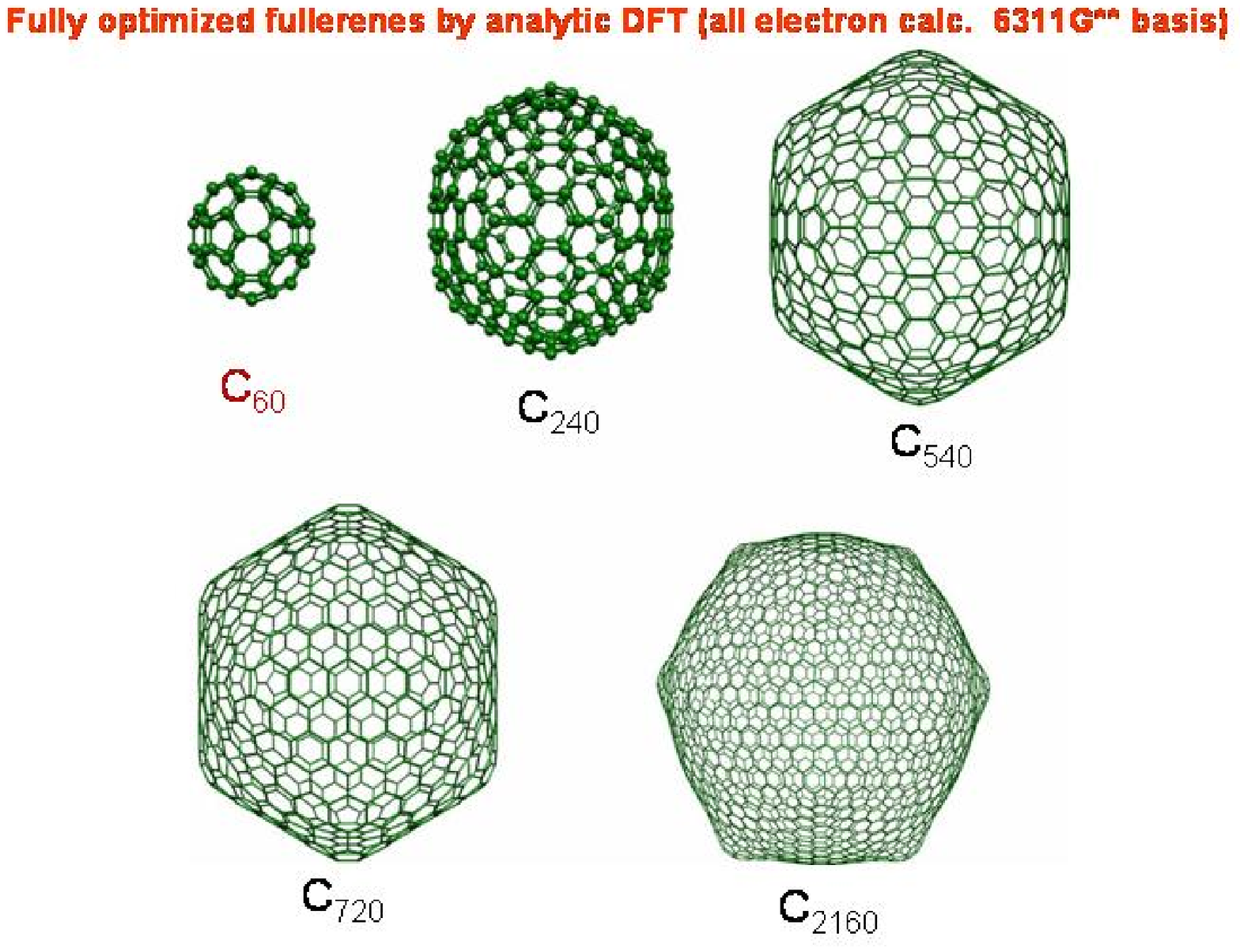,width=\linewidth,clip=true}
\caption{\label{fig:fullerenes} Fully optimized structures carbon fullerenes (Basis set: 6-311G**/Ahlrichs)
(see text for more details). }
\end{figure}

\begin{table}
\caption{ The median nearest-neighbor bond distance, mean radius, radial standard deviation, 
all in Angstroms, for the fullerenes of this work computed using  = 0.684667.  
The right-most column gives the atomization energy per atom (AE), in electron volts, that 
we compute using  = 0.64190.}
\label{table:F}
\begin{tabular}{llllrll}
\hline
Fullerene    & Median bond distance  &  Mean radius    &  AE \\

C$_{60}$    &	1.4244 &	3.5481  & -7.140 \\
C$_{240}$   &	1.4306 &	7.0728	& -7.373 \\
C$_{540}$   &	1.4264 &	10.5528	& -7.431 \\
C$_{960}$   &	1.4249 &	14.0342	& -7.459 \\
C$_{1500}$   &	1.4244 &	17.5225	& -7.474 \\
C$_{2160}$   &	1.4241 &	21.0137 &  -7.484 \\
\hline
\end{tabular}
\end{table}

\section{Carbon fullerenes}
      Carbon fullerenes structures larger than 100 atoms have been studied by 
several groups\cite{D91,Scuseria,Itoh}. Most of these 
studies  have used semiempirical models or tight biding methods or the Hartee-Fock theory 
plus minimal basis sets. 
Except for very recent calculation\cite{Calaminici} on \CTFT, fullerenes have not 
been studied using reasonable quality basis sets\cite{Scuseria2}.
This is principally because of large computational cost. We have used computationally 
efficient ADFT described above to optimize geometries of several carbon fullerenes from \CSX\, 
to \CTOS0\, using large polarized Gaussian basis sets of triple zeta quality
(6-311G**)\cite{ZD06}. 
The ADFT code developed in our group exploits  the icosahedral symmetry of these fullerenes in 
an efficient manner. Therefore, very large calculation on \CTOS0\, with about 39000 orbital 
basis functions, is still doable with modest computation resources.  
                  In order to get accurate geometries of larger fullerenes, we parametrize 
the ADFT to get the exact geometry of \CSX. This is accomplished by minimizing the mean square 
deviation between the experimental and predicted bond lengths of \CSX. This is possible without 
much difficulty as optimization of \CSX\, using  ADFT takes less than 5 minutes on single
processor Linux box (Intel(R) XEON(TM) CPU 2.20GHz with 2Gigabytes of Random Access Memory).
  The exact geometry of \CSX\, can be obtained for $\alpha=0.684667$.
We use this value of $\alpha$ for optimizing larger carbon fullerenes and hope that this 
will give accurate estimates of their geometries. The  \CNSXT\, fullerene can be optimized 
on single processor in 5 days.  The \CTOS0\, optimization was performed on Linux cluster using 
48 processors and took about 5 days.  The median bond distance of optimized carbon fullerenes 
is given in Table II and optimized structures are shown  in Fig. \ref{fig:fullerenes}.
We also made an attempt to get accurate atomization energies using the optimized geometries 
of fullerenes. For this purpose we reparametrize ADFT to get the exact binding energy of \CSX\, 
fullerene and use the $\alpha$ value thus determined to compute the atomization energies 
of larger fullerenes. These values are also given in Table II. However, such a procedure fails 
in that the atomization energy of \CTFT\, is already lower than that of graphite. Thus, to get
accurate estimate of atomization energy we need to go beyond the functional form that 
we have chosen in parameterizing the ADFT. Work is progress in our laboratory in this direction.

\section{conclusion}
     We have presented fully analytic implementation of density functional theory (ADFT). 
It uses analytic Gaussian basis sets to varitationally express the 
Kohn-Sham molecular orbitals, electron density and the one body effective potential 
of the density functional theory. The resultant formulation is 
computationally very efficient and allow for calulations on relatively large systems.
It permits use of atomic number dependent potential by means of Slater's exchange parameters.
Using the ADFT code, which efficiently uses the full point 
group symmetry of the molecule, we have optimized large inorganic fullerene-like cages and carbon
fullerenes containing more than two thousand atoms.

\end{document}